\documentclass[aps,prl,reprint,twocolumn,amsmath,amssymb]{revtex4-1}
\usepackage{stmaryrd}
\usepackage{times}
\usepackage{graphicx}
\usepackage{dcolumn}
\usepackage{bm}
\usepackage{color}
\begin{document}
\title{All-Electrical Generation of Spin-Polarized Currents in Quantum Spin Hall Insulators}
\author{L. L. Tao$^1$, K. T. Cheung$^1$, L. Zhang$^{2,1,3}$, and J. Wang$^1$}
\email[Electronic mail: ]{jianwang@hku.hk}
\affiliation{$^1$Department of Physics and the Center of Theoretical and Computational Physics, The University of Hong Kong, Hong Kong, China}
\affiliation{$^2$State Key Laboratory of Quantum Optics and Quantum Optics Devices, Institute of Laser Spectroscopy, Shanxi University, Taiyuan 030006, China}
\affiliation{$^3$Collaborative Innovation Center of Extreme Optics, Shanxi University, Taiyuan 030006, China}
\date{\today}
\begin{abstract}
The control and generation of spin-polarized currents (SPCs) without magnetic materials and external magnetic field is a big challenge in spintronics and normally requires spin-flip mechanism. In this work, we propose a novel method to control and generate SPCs in stanene nanoribbons in the quantum spin Hall (QSH) insulator regime by all electrical means without spin-flip mechanism. This is achieved with intrinsic spin-orbit coupling in stanene nanoribbons by tuning the relative phase of spin up and down electrons using a gate voltage, which creates a time delay between them thereby producing alternative SPCs driven by \emph{ac} voltage. The control and generation of SPCs are demonstrated numerically for \emph{ac} transport in both transient and \emph{ac} regime. Our results are robust against edge imperfections and generally valid for other QSH insulators such as silicene and germanene, \emph{etc}. These findings establish a novel route for generating SPCs by purely electrical means and open the door for new applications of semiconductor spintronics.
\end{abstract}
\maketitle
\emph{Introduction.} Spintronics exploits electron spin degrees of freedom which display many fascinating physics and bring many promising technological applications\cite{RevModPhys.76.323,nm409}. One of the central issues of spintronics is the control and generation of spin-polarized currents (SPCs). Recently, the generation and detection of SPCs at the nanoscale by purely electrical means has attracted great attention since this is the key step towards developing semiconductor spintronic devices\cite{science1359,nn759,nn35}. The conventional way of producing SPCs requires the application of magnetic fields\cite{nc1500,PhysRevB.92.035413} or ferromagnetic materials\cite{mrs389,ns44325}, which is conceptually different from the field-effect devices and difficult to integrate into the existing semiconductor devices\cite{np153}. Thus it would be highly desirable to generate SPCs by purely electrical means. Naturally, the system with spin-orbit coupling (SOC) in particular Rashba SOC is a promising candidate to achieve those objectives by exploiting the linking between an electron spin and its space motion\cite{nm871}. Indeed, it has been shown theoretically that SPCs can be produced by electrical means due to Rashba SOC in several systems, such as two-dimensional electron gas\cite{prb-changsheng}, graphene nanoribbons\cite{pccp16469}, and carbon nanotubes\cite{PhysRevB.93.165424}. Being a spin-flip SOC, Rashba SOC has been extensively investigated for generation of non-equilibrium spin polarization\cite{PhysRevB.78.212405,nn218}, spin-polarized current\cite{pccp16469,PhysRevB.93.165424} and pure spin current (spin Hall effect)\cite{RevModPhys.87.1213}. However, spin-flip SOC is detrimental to the spin lifetime due to the spin-flip scattering. New route to generate SPC without spin-flip SOC is clearly worth exploring. Since the generation of SPC normally requires spin-flip mechanism, a natural question arises: is it possible to induce a SPC using electrical means in the absence of spin-flip SOC?

Partial answers are available from the previous studies\cite{nn759,nn35,PhysRevB.74.155313,PhysRevB.81.115328}. It was demonstrated theoretically that opposite spin accumulations on the transverse edges can be induced by the strong lateral SOC (intrinsic SOC that conserves spin) in the two-dimensional electron gas\cite{PhysRevB.74.155313}, although a SPC can not be induced. Experimentally, Debray \emph{et al}.\cite{nn759} found a conductance plateau around half conductance quanta in a quantum point contact device with a large lateral SOC driven by asymmetry confining potential. However, it was found\cite{nn759,nn35,PhysRevB.81.115328} that in order to reproduce the plateau of half conductance quanta it is essential to consider a strong $e$-$e$ interaction acting like a spin-dependent potential, implying that magnetic-field-like term was required.

In this work, we propose a conceptually different way of generating SPCs using the intrinsic SOC, that conserves spin, in a series of quantum spin Hall (QSH) insulators, such as silicene\cite{PhysRevLett.108.155501}, germanene\cite{njp095002}, and stanene\cite{nm1020} whose band gap is very large (ranging $1$-$100$ meV)\cite{PhysRevB.84.195430,jpsj121003} in compared with graphene ($10^{-3}$ meV)\cite{PhysRevB.75.041401}. QSH insulators are a new state of quantum matter characterized by an insulating bulk and topologically protected gapless edge states\cite{PhysRevLett.95.226801,PhysRevLett.96.106802}. To produce SPCs using electrical means it seems that one must flip spin to make the transmission probabilities imbalance between spin up and down channels. The new working principle of our proposal is to tune the relative phase between the wave function of different spins that causes a time delay between spin up and down electrons traversing the system, which gives rise to SPCs in the \emph{ac} signals. Specifically, this is achieved by applying pulse or \emph{ac} source-drain voltage on the central-region-gated QSH insulator zigzag nanoribbons without any further decoration or doping. Without losing generality, we consider a practical QSH insulator device model, i.e., stanene zigzag nanoribbons as shown in Fig. \ref{f-1}. We emphasize that the analysis and discussion presented here are very general and are applicable to silicene and germanene as well. We also expect those findings to be a general feature of open quantum systems as long as a phase difference between spin up and down electrons can be established by electrical means.

\emph{The model and theoretical formalism.} The tight-binding Hamiltonian for describing the QSH insulator nanoribbon is given by\cite{PhysRevB.92.035413,jpsj121003,PhysRevB.89.195303},
\begin{equation}\label{f-1}
\begin{aligned}
    &H=-h_0\sum_{\langle i,j\rangle\alpha} c_{i\alpha}^{\dag}c_{j\alpha}+i\frac{\lambda_{SO}}{3\sqrt{3}}\sum_{\langle\langle i,j\rangle\rangle\alpha,\beta}\nu_{ij}c_{i\alpha}^{\dag}\sigma_z^{\alpha\beta}c_{j\beta}+\\
    &lE_z\sum_{i\alpha}\xi_ic_{i\alpha}^{\dag}c_{i\alpha}.
\end{aligned}
\end{equation}
The first term is the hopping term, where $c_{i\alpha}^{\dag}$ ($c_{j\alpha}$) is an electron creation (annihilation) operator at site $i$ ($j$) with spin $\alpha=\uparrow,\downarrow$, $h_0$ is the hopping energy and $\langle i,j\rangle$ denotes the sum over the nearest-neighbor sites. The second term represents the intrinsic SOC with strength $\lambda_{SO}$, $\langle\langle i,j\rangle\rangle$ denotes the sum over the next-nearest-neighbor sites, $\sigma_z$ is the $z$-component Pauli matrix and $\nu_{ij}=+1$ ($-1$) if the hopping is anticlockwise (clockwise) with respect to the $z$ axis\cite{PhysRevLett.95.226801}. The third term arises from the applied electric field $E_z$, $\xi_i=\pm1$ for $i=A, B$ site and $l$ is the buckling height. The following tight-binding parameters for stanane nanoribbon are used: $h_0=1.3$ eV, $\lambda_{SO}=0.1$ eV, $a=4.70$ {\AA}, and $l=0.40$ {\AA}\cite{jpsj121003}.

For the transient current calculation, the time-dependent terminal current $I_\alpha(t)$ of lead $\alpha$ is given by\cite{PhysRevB.87.205401}
\begin{equation}\label{eq-2}
I_\alpha(t)=2\text{ReTr}[\overline{\Gamma}_\alpha H_{cc}G^{<}_{cc}(t,t)\overline{\Gamma}_\alpha-i\overline{\Gamma}_\alpha\partial_tG^{<}_{cc}(t,t)\overline{\Gamma}_\alpha],
\end{equation}
where $\overline{\Gamma}_\alpha$ is an auxiliary projection matrix, $G^{<}_{cc}(t,t)$ and $H_{cc}$ are the time-dependent lesser Green's function and the Hamiltonian of the central scattering region, respectively. We consider both upward and downward voltage pulse\cite{PhysRevB.87.205401,order1}. In \emph{ac} regime, \emph{ac} current consists of particle current and displacement current\cite{PhysRevLett.70.4114,PhysRevLett.82.398,PhysRevB.79.195315}, namely the total dynamic conductance $G_{LR}=G^c_{LR}+G^d_{LR}$. Particle conductance $G_{LR}^{c}$ can be expressed as\cite{PhysRevLett.70.4114}
\begin{equation}\label{eq-3}
    G_{LR}^{c}=-\int\frac{dE}{2\pi}\frac{f-\bar{f}}{\omega}\text{Tr}[s_{LR}^\dag(E) s_{LR}(E+\omega)],
\end{equation}
where $f$ is the Fermi function, $s_{LR}$ the scattering matrix, $\omega$ photon frequency and $\bar{f}\equiv f(E+\omega)$. The complicated expression of displacement conductance $G^d_{LR}$ is given in the Supplemental Material\cite{sm}.
\begin{figure}
\includegraphics[width=0.45\textwidth]{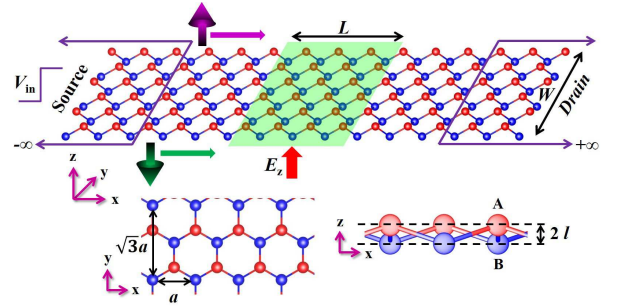}%
\caption{\label{f-1} (a) Schematic plot of a zigzag nanoribbon device with different views. Here $E_z$ is the external electric field generated by the back-gate voltage. The source and drain electrodes extend to $x=\mp\infty$ as shown by the left/right pointing arrows. $V_{\text{in}}$ is the input pulse voltage. $a$ is the lattice constant. $l$ is the buckling height. $W$ (unit: $\sqrt{3}a$) and $L$ (unit: $a$) denote the width and the length of the gated region, respectively.}
\end{figure}

\begin{figure}
\includegraphics[width=0.45\textwidth]{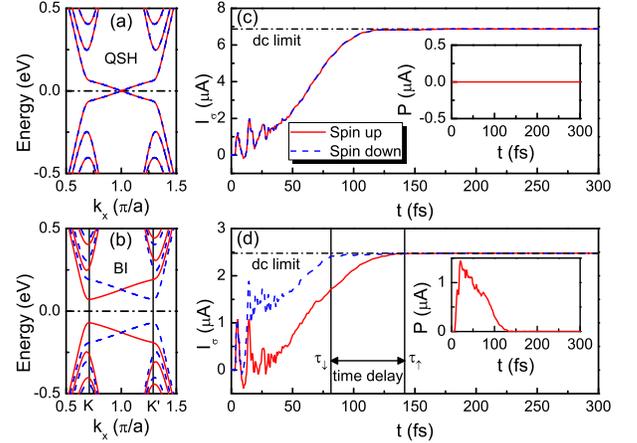}%
\caption{\label{f-2} (a-b) Band structures for the nanoribbon with $W=12$. (a) QSH insulator phase with $lE_z=0$. (b) trivial BI phase with $lE_z=0.1h_0$. $K$ ($0.71$) and $K'$ ($1.29$) denote the position of conduction band minimum (or valence band maximum) for spin up and down, respectively. (c-d) The transient current $I_\sigma(t)$ ($\sigma=\uparrow$ for spin up, $\downarrow$ for spin down) through the nanoribbon in response to a upward step pulse. $W=12$, $L=16$, and $V_b(t) =0.2 \theta(t) $ V. (c) $lE_z=0$ and (d) $lE_z=0.1h_0$. Inset: the spin-polarized current defined as: $P(t)=I_\downarrow(t)-I_\uparrow(t)$. $\tau_{\uparrow}$ ($\tau_{\downarrow}$) denotes roughly the characteristic time for spin up (down) electron reaching the steady state. The short-dash lines that denote the \emph{dc} current for spin up/down are calculated by using the Landauer-B\"uttiker formula.}
\end{figure}

\emph{Results and discussion.} We first analyze the electronic structure of the stanene zigzag nanoribbon. Previous work show that the stanene as well as silicene and germanene reveal rich topological phases driven by $E_z$\cite{nc1500,jpsj121003}, which is distinct from graphene. Fig. \ref{f-2}(a-b) show the band structures for the nanoribbon with and without $E_z$. In the absence of $E_z$ shown in Fig. \ref{f-2}(a), the nanoribbon is in the QSH insulator phase, characterized by the gapless edge states and the band is spin degenerate. For a finite $E_z$ larger than the critical field shown in Fig. \ref{f-2}(b), a finite gap is opened up and the nanoribbon is driven into a trivial band insulator (BI) phase. More interestingly, the spin degeneracy is lifted due to the combined effect of the inversion symmetry breaking by $E_z$ and the intrinsic SOC, which is the key for the generation of SPCs. Note that with $E_z$ turned on the spin component $s_z$ is still a good quantum number, as evident from $[s_z, H]=0$.

We now connect our system with two leads consisting of stanene nanoribbon and study the transport behaviors. Fig. \ref{f-2}(c-d) shows the spin-resolved transient current $I_\sigma(t)$ and spin-polarized current $P(t)=I_\downarrow(t)-I_\uparrow(t)$ for an upward voltage pulse. Overall, $I_\sigma(t)$ evolves in three distinct regimes: initial strong oscillating regime, followed by steady increasing and finally approaching to the \emph{dc} steady state. For the central region in the QSH insulator phase [see Fig. \ref{f-2}(c)], the transient current is spin degenerate at all times. In contrast, for the central region in the BI phase induced by $E_z$ [see Fig. \ref{f-2}(d)], the current is strongly spin-polarized before reaching the steady state, as shown explicitly in the inset. Therefore, the generation of SPC can be realized in the time window away from the steady state. Interestingly, the characteristic time reaching the steady state is clearly different for spin up and down electrons, namely $\tau_{\uparrow}>\tau_{\downarrow}$, showing a time delay between spin up and down electrons. As we shall see latter, this time delay is due to the \emph{generic} phase difference between spin up and down electrons traversing the system. Inset of Fig. \ref{f-2}(d) shows that $P$ reaches the maximum at certain point in the transient regime and the time window for SPC is about $130$ fs. We also calculated the transient current for the nanoribbons with different lengths $L$ (barrier thickness) as shown in Fig. S3(a). Qualitatively, they all display fairly similar features as compared with the case of $L=16$. Moreover, the characteristic time $\tau_{\uparrow}$ ($\tau_{\downarrow}$) scales almost linearly with the barrier thickness [see Fig. S3(b)], however their slopes are different, which results in the linearly increase of time delay with barrier thickness. The pulse amplitude dependency of SPC is given in Fig. S4(b). It is seen that the magnitude of SPC increases monotonously with increasing pulse amplitude, while the time window producing the SPC does not change much. On the other hand, as shown in Fig. S2, we confirmed numerically that the SPC also occurs in the transient regime for a downward voltage pulse. This indicates that SPC can be generated for a periodic train of pulses\cite{note1}. Note that previous work demonstrated that a SPC can be produced from the time-dependent Rashba SOC generated by a time-dependent gate voltage\cite{PhysRevB.68.233307,PhysRevB.71.195314,PhysRevB.78.245312}. Its physical mechanism generating SPC is distinctly different from our work, in which the Rashba SOC is absent.

\begin{figure}
\includegraphics[width=0.45\textwidth]{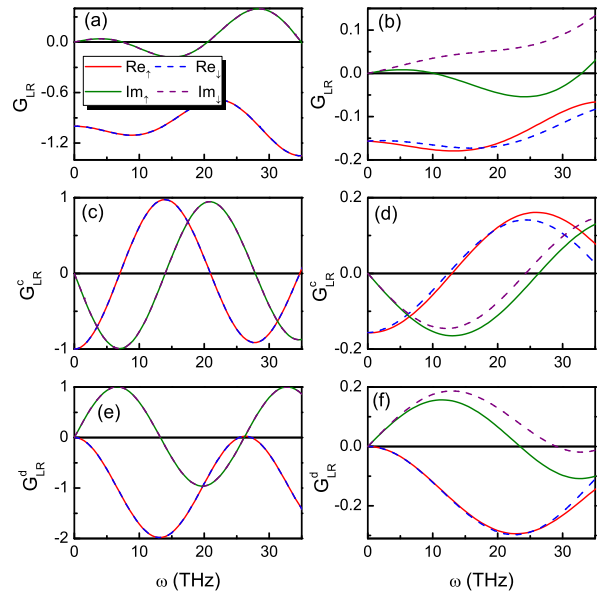}%
\caption{\label{f-3} Dynamic conductance (in units of $e^2/h$) as a function of frequency $\omega$. (a, b) The total dynamic conductance $G_{LR}$. (c, d) The dynamic conductance $G_{LR}^{c}$ contributed by the particle current. (e, f) The dynamic conductance $G_{LR}^{d}$ contributed by the displacement current. (a, c, e) and (b, d, f) show the results for the nanoribbon without and with $E_z$ in the central region, respectively. Re: real part, Im: imaginary part. Parameters $W$, $L$, and $E_z$ are the same as those given in the caption of Fig. \ref{f-2}.}
\end{figure}
\begin{figure}
\includegraphics[width=0.45\textwidth]{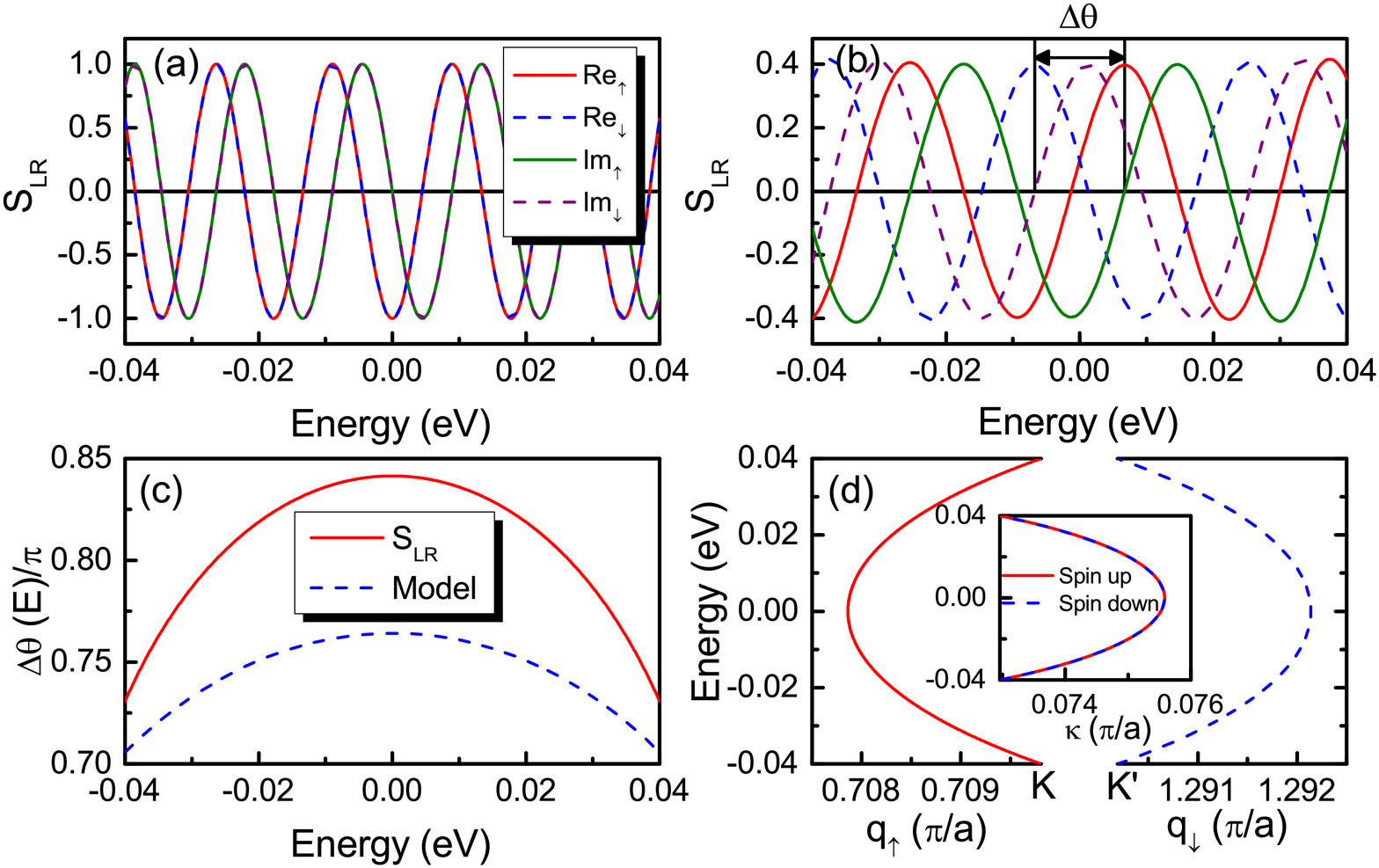}%
\caption{\label{f-4} The scattering matrix $s_{LR}$ as a function of energy for (a) QSH insulator phase and (b) BI phase in the central region. (c) The phase difference $\Delta\theta=\theta_{\downarrow}-\theta_{\uparrow}$ as a function of energy when the central region is in the BI phase. $S_{LR}$: $\Delta\theta(E)$ is determined from scattering matrix. Model: $\Delta\theta(E)$ is estimated from $\Delta\theta(E)=[q_\downarrow(E)-q_\uparrow(E)]d$. Parameters $W$, $L$, and $E_z$ are the same as those given in the caption of Fig. \ref{f-2}. (d) The complex band structure for the nanoribbon in the BI phase. $q$ and $\kappa$ (inset) correspond to the real and imaginary components of wave vector $k_x=q+i\kappa$, respectively.}
\end{figure}
To understand the physics behind the generation of SPC in the transient regime, we study the dynamical conductance $G_{\alpha\beta}(\omega)$ ($\alpha,\beta=L, R$) as a function of frequency $\omega$ for a sinusoidal bias in the linear response regime since a train of bias pulse can be expanded in terms of simple harmonics with basic frequency. In Fig. \ref{f-3}, we show the frequency dependency of $G_{LR}(\omega)$. For the central region in the QSH insulator phase shown in Fig. \ref{f-3}(a), $G_{LR}(\omega)$ are spin degenerate and exhibit oscillation features due to the interplay of the particle current and displacement current, as evident from Fig. \ref{f-3}(c, e). The particle current $G_{LR}^{c}$ is found to be $G_{LR}^{c}(\omega)=-e^{i2\pi\frac{\omega}{E_0}}$ with $E_0=17.8$ meV which can be understood from the scattering matrix $s_{LR}(E)$ in Fig. \ref{f-4}(a) which relates $G^c_{LR}$ through Eq. \ref{eq-3}. Analytically, we find $s_{LR}(E)=-e^{i2\pi\frac{E}{E_0}}$ near the Fermi energy, $E_0=2\pi v_F/d$\cite{sm}, where $v_F$ is the Fermi velocity of edge states and $d$ is the realistic thickness\cite{note2}. Note that the energy dependence of $s_{LR}$ is the generic feature of massless Dirac fermion. Using $v_{F}$ determined from the band dispersion in Fig. \ref{f-2}(b), we find $E_0=17.3$ meV which agrees well with numerical result $E_0=17.8$ meV.

For the central region in the BI phase shown in Fig. \ref{f-3}(b, d, f), the most striking observation is that the spin degeneracy of dynamic conductance is lifted. The spin-splitting $G_{LR}^{c}$ can be understood from $s_{LR}(E)$ as shown in Fig. \ref{f-4}(b). It is clearly seen that there is a phase difference $\Delta\theta(E)$ for $s_{LR}(E)$ between spin up and down electrons. Importantly this phase difference has to be energy dependent in order to lift the spin degeneracy as seen from Eq. \ref{eq-3}. From Fig. \ref{f-4}(c), we find that $\Delta\theta(E)$ is indeed weakly energy dependent, which is responsible for spin-splitting $G_{LR}^{c}$.

The observed $\Delta\theta(E)$ can be understood from the physics of one dimensional quantum tunneling. In contrast to the conventional quantum tunneling picture where the incident electron decays exponentially inside the barrier, the incident edge state (Dirac fermion) near the Fermi energy decays in an oscillatory fashion as $e^{-\kappa x}e^{iq x}$ through the barrier (the gap region) so that an extra phase $qd$ is acquired after the tunneling. This unusual feature is confirmed from the complex band structure as shown in Fig. \ref{f-4}(d), from which the complex wave vector $k_x=q+i\kappa$ can be determined for spin up and down electrons. This behavior of oscillatory decay of edge/surface states was previously reported in the QSH insulator of Bernevig, Hughes, and Zhang model\cite{PhysRevB.90.155307,science1757} and also in the 3D topological insulator Bi$_2$Se$_3$\cite{jpcm395501}. Note that the position of conduction band minimum (or valence band maximum) [see Fig. \ref{f-2}(b)] for spin up (down) electron is located near $K$ ($K'$) point giving rise to a different $q_\uparrow$ or $q_\downarrow$ while $\kappa$ is spin degenerate due to the time reversal symmetry [inset of Fig. \ref{f-4}(d)]. In other words, the electron with different spins acquire different phases after the tunneling with the phase difference $\Delta\theta(E)=[q_\downarrow(E)-q_\uparrow(E)]d$, where $q_\downarrow$ and $q_\uparrow$ are around the $K'$ and $K$ point, respectively. The comparison between the model and numerical results is shown in Fig. \ref{f-4}(c). We see that our simple one dimensional model captured essential physics. Since the transmission coefficient in the first subband $T(E)$ depends weakly on energy [see Fig. S1(b)], the phase accumulated after the tunneling is related to the tunneling time defined as $\tau_t \equiv h {\rm Im} (s_{LR} \partial_E s^\dagger_{LR})/T(E)=h \partial_E \theta(E)$\cite{rmp}, suggesting a time delay $t_0=h \partial_E \Delta \theta(E)=h(\partial q_\uparrow -\partial q_\downarrow) d$ between spin up and down electrons. Indeed, from Fig. S3(b), we see that the time delay $t_0$ increases linearly with the barrier thickness. Now the physics becomes transparent. Spin up electron passing through the energy gap of BI region experiences a time delay relative to spin down electron. While this does not change the spin polarization of \emph{dc} current since it is time independent, the spin polarization of any \emph{ac} transport properties will be affected as shown above. We conclude that the SPC arises from the combined effect of $E_z$ and \emph{ac} field: (1) $E_z$ induces an energy dependent phase difference between spin up and down electrons resulting a time delay between them; (2) this time delay can only be manifested through the inelastic scattering (integrand of Eq. \ref{eq-3}) due to \emph{ac} field.

It is known that the local defects in particular the atomic vacancy are usually existed at the edges of the 2D honeycomb structures\cite{PhysRevB.73.085421,PhysRevB.73.125415}. To explore the effect of vacancy, we calculated the transient current for the nanorribbon with different edge-vacancy configurations. From Fig. S5(a), the edge vacancies modify slightly the magnitude of the currents in the steady state (\emph{dc}-limit), while the induced SPCs in the transient regime do not change much. In addition, these vacancies have little effect on the conductance especially in the lower subbands [see Fig. S5(b)].

\emph{Conclusion.} We have proposed a novel way of generating a SPC in a family of QSH insulators, such as stanene, silicene, and germanene \emph{etc.} zigzag nanoribbons using electrical means. Conceptually different from the conventional all-electrical approaches in producing a SPC, the spin-flip SOC is not required here. The SPC is generated by tuning the phase difference between spin up and down electrons induced by a gate voltage. Since the bulk gap of stanene was predicted to be sufficiently large for practical applications at room temperature\cite{PhysRevLett.111.136804}, our proposal for the generation of SPCs in QSH insulators by purely electrical means is achievable in experiment.

\emph{Acknowledgement.} L.L.T. would like to thank Zhizhou Yu and Yanxia Xing for useful discussions. This work is supported by University Grant Council of Hong Kong (Contract No. AoE/P-04/08) and the National Natural Science Foundation of China (Grant No. 11374246).


\begin{thebibliography}{99}
\bibitem{RevModPhys.76.323} I. \ifmmode \check{Z}\else \v{Z}\fi{}uti\ifmmode \acute{c}\else \'{c}\fi{}, J. Fabian, and S. Das Sarma, Rev. Mod. Phys. {\bf 76}, 323 (2004).
\bibitem{nm409} D. Pesin and A. H. MacDonald, Nat. Mater. {\bf 11}, 409 (2012).
\bibitem{science1359} R. Crook, J. Prance, K. J. Thomas, S. J. Chorley, I. Farrer, D. A. Ritchie, M. Pepper, and C. G. Smith, Science {\bf 312}, 1359 (2006).
\bibitem{nn759} P. Debray, S. M. S. Rahman, J. Wan, R. S. Newrock, M. Cahay, A. T. Ngo, S. E. Ulloa, S. T. Herbert, M. Muhammad, and M. Johnson, Nat. Nanotechol. {\bf 4}, 759 (2009).
\bibitem{nn35} P. Chuang, S. C. Ho, L. W. Smith, F. Sfigakis, M. Pepper, C. H. Chen, J. C. Fan, J. P. Griffiths, I. Farrer,	H. E. Beere, G. A. C. Jones, D. A. Ritchie, and T. M. Chen, Nat. Nanotechol. {\bf 10}, 35 (2015).
\bibitem{nc1500} W. F. Tsai, C. Y. Huang, T. R. Chang, H. Lin, H. T. Jeng, and A. Bansil, Nat. Commun. {\bf 4}, 1500 (2013).
\bibitem{PhysRevB.92.035413} Kh. Shakouri, H. Simchi, M. Esmaeilzadeh, H. Mazidabadi, and F. M. Peeters, Phys. Rev. B {\bf 92}, 035413 (2015).
\bibitem{mrs389} S. Parkin, MRS Bull. {\bf 31}, 389 (2006).
\bibitem{ns44325} J. Tang and K. L. Wang, Nanoscale {\bf 7}, 4325 (2015).
\bibitem{np153} D. D. Awschalom and M. E. Flatte, Nat. Phys. {\bf 3}, 153 (2007).
\bibitem{nm871} A. Manchon, H. C. Koo,	J. Nitta, S. M. Frolov, and R. A. Duine, Nat. Mater. {\bf 14}, 871 (2015).
\bibitem{prb-changsheng} C. S. Li, Y. J. Yu, Y. D. Wei, and J. Wang, Phys. Rev. B {\bf 75}, 035312 (2007).
\bibitem{pccp16469} L. Chico, A. Latg\'e, and L. Brey, Phys. Chem. Chem. Phys. {\bf 17}, 16469 (2015).
\bibitem{PhysRevB.93.165424} H. Santos, A. Latg\'e, J. E. Alvarellos, and L. Chico, Phys. Rev. B {\bf 93}, 165424 (2016).
\bibitem{PhysRevB.78.212405} A. Manchon and S. Zhang, Phys. Rev. B {\bf 78}, 212405 (2008).
\bibitem{nn218} C. H. Li, O. M. J. vant Erve, J. T. Robinson, Y. Liu, L. Li, and B. T. Jonker, Nat. Nanotechol. {\bf 9}, 218 (2014).
\bibitem{RevModPhys.87.1213} J. Sinova, S. O. Valenzuela, J. Wunderlich, C. H. Back, and T. Jungwirth, Rev. Mod. Phys. {\bf 87}, 1213 (2015).
\bibitem{PhysRevLett.94.246601} F. Zhai and H. Q. Xu, Phys. Rev. Lett. {\bf 94}, 246601, (2005).
\bibitem{PhysRevB.74.155313} Y. X. Xing, Q. F. Sun, L. Tang, and J. P. Hu, Phys. Rev. B {\bf 74}, 155313 (2006).
\bibitem{PhysRevB.81.115328} A. T. Ngo, P. Debray, and S. E. Ulloa, Phys. Rev. B {\bf 81}, 115328 (2010).
\bibitem{PhysRevLett.108.155501} P. Vogt, P. D. Padova, C. Quaresima, J. Avila, E. Frantzeskakis, M. C. Asensio, A. Resta, B. Ealet, and G. L. Lay, Phys. Rev. Lett. {\bf 108}, 155501 (2012).
\bibitem{njp095002} M. E. D\'avila, L. Xian, S. Cahangirov, A. Rubio, and G. Le Lay, New J. Phys. {\bf 16}, 095002 (2014).
\bibitem{nm1020} F. Zhu, W. Chen, Y. Xu, C. Gao, D. Guan, C. Liu, D. Qian, S. C. Zhang, J. Jia, Nat. Mater. {\bf 14}, 1020 (2015).
\bibitem{PhysRevB.84.195430} C. C. Liu, H. Jiang, and Y. Yao, Phys. Rev. B {\bf 84}, 195430 (2011).
\bibitem{jpsj121003} M. Ezawa, J. Phys. Soc. Jpn. {\bf 84}, 121003 (2015).
\bibitem{PhysRevB.75.041401} Y. Yao, F. Ye, X. L. Qi, S. C. Zhang, and Z. Fang, Phys. Rev. B {\bf 75}, 041401 (2007).
\bibitem{PhysRevLett.95.226801} C. L. Kane and E. J. Mele, Phys. Rev. Lett. {\bf 95}, 226801 (2005).
\bibitem{PhysRevLett.96.106802} B. A. Bernevig and S. C. Zhang, Phys. Rev. Lett. {\bf 96}, 106802 (2006).
\bibitem{PhysRevLett.107.076802} C. C. Liu, W. Feng, and Y. Yao, Phys. Rev. Lett. {\bf 107}, 076802 (2011).
\bibitem{PhysRevLett.111.136804} Y. Xu, B. Yan, H. J. Zhang, J. Wang, G. Xu, P. Tang, W. Duan, and S. C. Zhang, Phys. Rev. Lett. {\bf 111}, 136804 (2013).
\bibitem{PhysRevB.89.195303} S. Rachel and M. Ezawa, Phys. Rev. B {\bf 89}, 195303 (2014).
\bibitem{PhysRevB.87.205401} L. Zhang, J. Chen, and J. Wang, Phys. Rev. B {\bf 87}, 205401 (2013).
\bibitem{order1} K. T. Cheung, B. Fu, Z. Yu, and J. Wang, arXiv:1602.01638.
\bibitem{PhysRevLett.70.4114} M. B\"uttiker, A. Pr\^etre, and H. Thomas, Phys. Rev. Lett. {\bf 70}, 4114 (1993).
\bibitem{PhysRevLett.82.398} B. Wang, J. Wang, and H. Guo, Phys. Rev. Lett. {\bf 82}, 398 (1999).
\bibitem{PhysRevB.79.195315} Y. Wei and J. Wang, Phys. Rev. B {\bf 79}, 195315 (2009).
\bibitem{sm} See Supplemental Material for the resluts of transient current in response to a downward pulse, barrier length and pulse amplitude dependent transient currents, effect of edge vacancies on the transient currents, and technical details of the CAP method, transient current formalism, complex band, scattering matrix, and dynamic conductance calculation, which includes Refs.\cite{PhysRevB.87.205401,order1,PhysRevLett.82.398,PhysRevB.79.195315}.
\bibitem{note1} Note that the transient current under a periodic voltage pulse is not studied in this work. Consequently, the effect of pulse shape and duration cannot be investigated directly. However, the essential physics discussed in this work are not affected by the pulse shape or duration.
\bibitem{PhysRevB.68.233307} A. G. Mal'shukov, C. S. Tang, C. S. Chu, and K. A. Chao, Phys. Rev. B {\bf 68}, 233307 (2003).
\bibitem{PhysRevB.71.195314} C. S. Tang, A. G. Mal'shukov, and K. A. Chao, Phys. Rev. B {\bf 71}, 195314 (2005).
\bibitem{PhysRevB.78.245312} C. H. Lin, C. S. Tang, and Y. C. Chang, Phys. Rev. B {\bf 78}, 245312 (2008).
\bibitem{note2} In our calculations, $L$ denotes the number of layers being gated by $E_z$. The realistic thickness $d$ (unit: $a$) of the device is $d=L+12-1$ (the buffer layer is fixed as $6$) for the central layer in the QSH insulator phase, while it is $d=L-1$ for the central layer in the BI phase.
\bibitem{PhysRevB.90.155307} X. Dang, J. D. Burton, A. Kalitsov, J. P. Velev, and E. Y. Tsymbal, Phys. Rev. B {\bf 90}, 155307 (2014).
\bibitem{science1757} B. A. Bernevig, T. L. Hughes, S. C. Zhang, Science {\bf 314}, 1757 (2006).
\bibitem{jpcm395501} J. Betancourt, S. Li, X. Dang, J. D. Burton, E. Y. Tsymbal and J. P. Velev, J. Phys.: Condens. Matter {\bf 28}, 395501 (2016).
\bibitem{rmp} E. H. Hauge and J. A. Stovneng, Rev. Mod. Phys. {\bf 61}, 917 (1989).
\bibitem{PhysRevB.73.085421} Y. Niimi, T. Matsui, H. Kambara, K. Tagami, M. Tsukada, and H. Fukuyama, Phys. Rev. B {\bf 73}, 085421 (2006).
\bibitem{PhysRevB.73.125415} Y. Kobayashi, K. Fukui, T. Enoki, and K. Kusakabe, Phys. Rev. B {\bf 73}, 125415 (2006).
\end{thebibliography}
\end{document}